\documentclass[12pt]{iopart}

\newcommand{\csquare}{\mathbin{\mathchoice
  {\xsquare\scriptstyle}
  {\xsquare\scriptstyle}
  {\xsquare\scriptscriptstyle}
  {\xsquare\scriptscriptstyle}
}}
\newcommand{\xsquare}[1]{\vcenter{\hbox{$#1\square$}}}

\newcommand{\cvartriangle}{\mathbin{\mathchoice
  {\xvartriangle\scriptstyle}
  {\xvartriangle\scriptstyle}
  {\xvartriangle\scriptscriptstyle}
  {\xvartriangle\scriptscriptstyle}
}}
\newcommand{\xvartriangle}[1]{\vcenter{\hbox{$#1\vartriangle$}}}

\def\bra#1{\langle{#1}|}
\def\ket#1{|{#1}\rangle}

{\catcode`\|=\active 
  \gdef\Braket#1{\begingroup
\mathcode`\|32768\let|\BraVert\left<{#1}\right>\endgroup}}
\def\BraVert{\egroup\,\mid\,\bgroup}

\expandafter\let\csname equation*\endcsname\relax
  \expandafter\let\csname endequation*\endcsname\relax
\usepackage{amsmath, amsfonts, amssymb,mathalfa,dsfont}
    \usepackage{graphicx}
    \usepackage{hyperref}
    \usepackage{color, soul}
\DeclareMathOperator{\Span}{span}

\allowdisplaybreaks

\begin{document}

\title{Energy diffusion in the ergodic phase of a many body localizable spin chain}

\author{V.\ K.\ Varma}
\address{The Abdus Salam ICTP, Strada Costiera 11, 34151, Trieste, Italy}

\author{A.\ Lerose}
\address{The Abdus Salam ICTP, Strada Costiera 11, 34151, Trieste, Italy}
\address{Department of Physics, Universit\`a degli Studi di Milano, Milano, Italy}
\address{SISSA-International School for Advanced Studies, via Bonomea, 265, 34136 Trieste, Italy}

\author{F.\ Pietracaprina}
\address{Dipartimento di Fisica, La Sapienza Universit\`a di Roma, 00185 Roma, Italy}
\address{SISSA-International School for Advanced Studies, via Bonomea, 265, 34136 Trieste, Italy}

\author{J.\ Goold}
\address{The Abdus Salam ICTP, Strada Costiera 11, 34151, Trieste, Italy}

\author{A.\ Scardicchio}
\address{The Abdus Salam ICTP, Strada Costiera 11, 34151, Trieste, Italy}
\address{INFN, Sezione di Trieste, Via Valerio 2, 34126, Trieste, Italy}
\address{Dipartimento di Fisica, Universit\`a degli Studi di Bari ``Aldo Moro'', 70126, Bari, Italy}

\date{\today}

\begin{abstract}
The phenomenon of many-body localization in disordered quantum many-body systems occurs when all transport is suppressed despite the excitations of the system being interacting. 
In this work we report on the numerical simulation of autonomous quantum dynamics for disordered Heisenberg chains when the system is prepared with an initial 
inhomogeneity in the energy density profile. Using exact diagonalisation and a dynamical code based on Krylov subspaces we are able to simulate dynamics for up to $L=26$ spins. We find, 
surprisingly, the breakdown of energy diffusion even before the many-body localization transition whilst the system is still in the ergodic phase. 
Moreover, in the ergodic phase we also find a large region in parameter space where the energy dynamics remains diffusive but where spin transport has been previously evidenced to  
occur only subdiffusively: this is found to be true for initial states composed of infinitely many hydrodynamic modes (square-wave energy profile) or just the single longest mode (sinusoidal profile).
This suggestive finding points towards a peculiar ergodic phase where particles are transported slower than energy, reminiscent of the situation in amorphous solids and of the gapped phase of the 
anisotropic Heisenberg model. 
\end{abstract} 

\maketitle

\makeatletter

\section{Introduction}
The theory of disordered quantum systems aims to understand transport in a wide range of paradigms in condensed matter physics. 
This is due in large part to the seminal work of 
Anderson in 1958 \cite{Anderson:58} who found that sufficiently strong disorder was enough to completely localize an electron on a disordered lattice leading to the absence of 
diffusion. 
The lack of transport for sufficiently strong disorder, and its absence in one and two dimensions remains under intense investigation since its 
original inception \cite{Abrahams:10,Lee:85}. 

The phenomenon of many-body localization (MBL) is the persistence of localization, and hence complete suppression of transport, even in the presence of interactions: an initial macroscopic inhomogeneity in the energy density profile of the system persists over arbitrarily long times. 
The possibility that the localized phase could be stable to weak interactions was first put forward in the seminal work of Basko, Aleiner 
and Altshuler (BAA) \cite{Basko:06} and was surprising given that the general consensus was that interactions should lead to collisional dephasing and hence 
delocalization. 
Following this impetus, the past decade has seen a surge of studies related to the properties of the MBL phase \cite{Oganesyan:07,Luitz:15,Lev:15}, its rich phenomenology \cite{husereview,altmanreview}, 
including emergent integrability \cite{imbrie,ros15,chandran15}, various approximation methods to numerically analyse its properties 
\cite{pekker,khemani,voskprx,potterprx,Pietracaprina:15a}, and the presence of a many-body mobility edge 
\cite{Laumann:14a,Baldwin:15a,Luitz:15,abaninmob:15}.

The ergodic phase, however, has received less attention, firstly because the numerics is more demanding and secondly the expectation is that the ergodic phase is generic and 
hence less ``interesting" than the newly discovered MBL phase. However recent works have pointed towards a highly nontrivial ergodic region. 
In the ergodic region of MBL Hamiltonians the entanglement dynamics is characterized by a power-law growth of entanglement entropy \cite{Chiara:06,Znidaric:08,Bardarson:12,Kjall:14} 
(contrasting with logarithmic growth in the localized region \cite{Bardarson:12,Vosk:13,Abanin:13,Vosk:14,Nanduri:14}). 
In terms of the transport of conserved quantities, evidence has mounted for a regime of subdiffusion for the spin transport in the ergodic phase \cite{Agarwal,Lev:15,Znidaric:16}, as well as the presence of 
Griffiths rare regions close to the transition leading to anomalous power laws in certain spectral functions \cite{Gopalakrishnan, Agarwal,potterprx}. 
Kim and Huse have previously demonstrated, in an ergodic but not disordered spin chain, that the 
entanglement grows linearly with time while the energy is transported diffusively \cite{Huse:13}, suggesting a relation between entanglement transport and energy transport \cite{potterprx,voskprx}. 

In this study, we provide evidence that \emph{there is a considerable portion of the ergodic phase where energy transport is diffusive, but the diffusive behaviour breaks down well 
before the many-body localisation transition}. 
Moreover earlier works have argued for anomalous spin transport in the ergodic phase \cite{Lev:15, Agarwal, Luitz:15a, Znidaric:16}; 
it therefore behoves us to ask how energy is transported in such systems, in particular whether it is at the same or different rate as the spin transport. 
The latter scenario, where energy and spin excitations are transported at 
different rates, can occur if they are decoupled or weakly coupled leading to different transit times across the insulating or critical Griffiths islands, with the effect being exponentially exacerbated 
in the thermodynamic limit \cite{Gopalakrishnan:15}. Our numerics points towards this scenario.

These two findings taken together point towards a highly nontrivial ergodic region where (i) energy stops diffusing well before the localized phase, 
and (ii) energy does diffuse in parts of it but particles are transported slower, reminiscent of the situation in amorphous solids (glasses) \cite{glasses} and the gapped phase of the anisotropic Heisenberg chain \cite{Znidaric:11, Meisner}. 
This could be related to the possible existence of a so-called 
\emph{non-ergodic extended} phase in the Anderson problem on the Bethe lattice or in high dimension \cite{MultiFractal, Kamenev, goold, intermediate}. 
Moreover, comparing to the results of the recent work \cite{Luitz:15a}, we find that the energy diffusion breaks down very close to the point when the entropy spread becomes subdiffusive, 
a manifestation of the entanglement entropy growth dominating the growth of correlation functions of physical observables. 

\begin{figure}[htb]
\centering
  \begin{tabular}{@{}ll@{}}
    \hspace*{+0.0cm}\raisebox{-0.03ex}{\includegraphics[width=.45\textwidth,height=4.8cm]{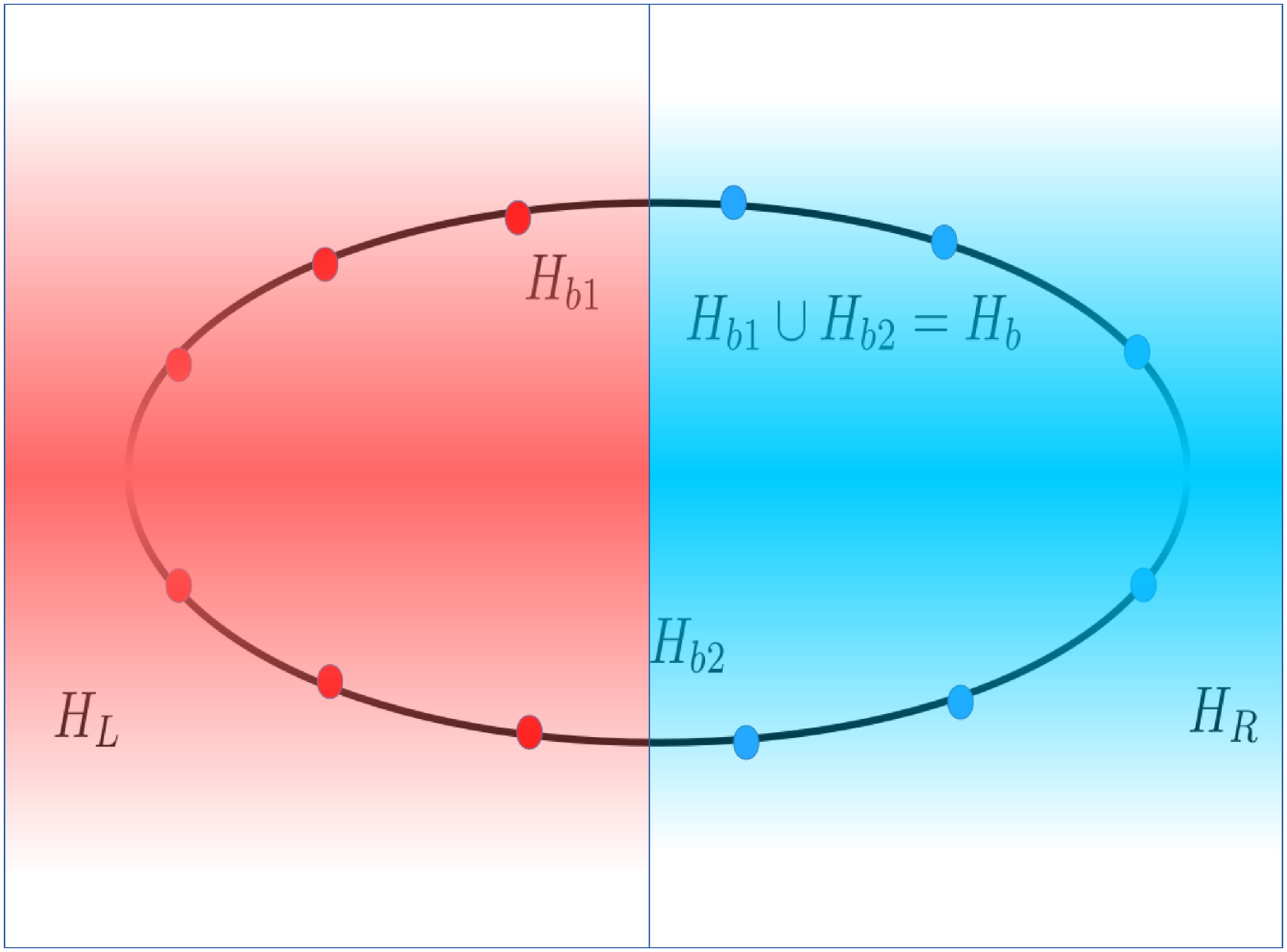}} &
    \hspace*{-0.cm}\raisebox{-0.03ex}{\includegraphics[width=.45\textwidth,height=4.8cm]{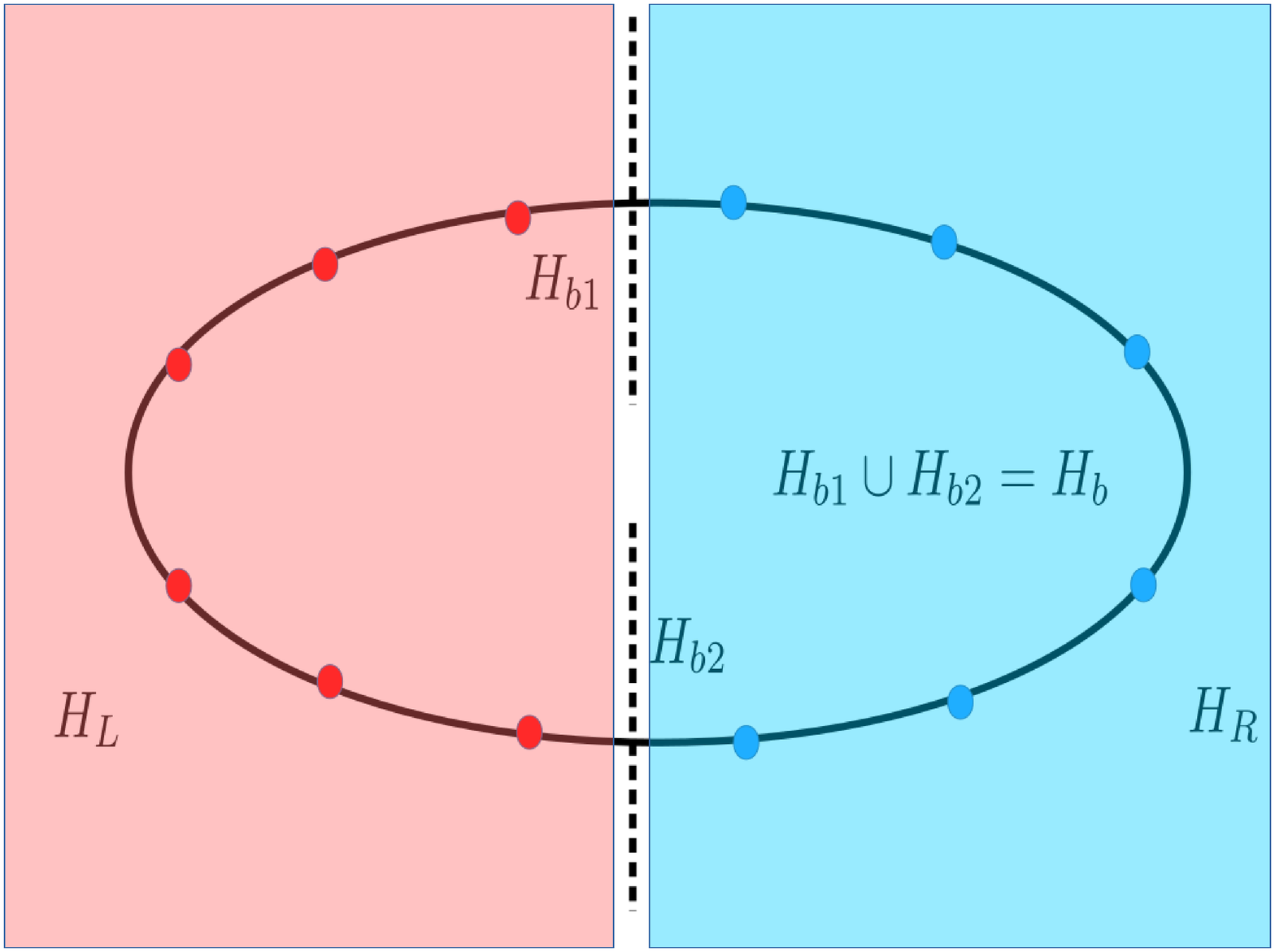}} 
  \end{tabular}
  \caption{(Colour online) Partition of the disordered chain into hot (red) and cold (blue) regions at time $t=0$. 
  In the first set-up (left) the hot and cold regions are inhomogeneously distributed across the chain's halves as a sinusoidal wave, with the hotness/coldness being maximal at the 
  centre of the right/left halves and the energy density being smoothly connected (indicated by white) at the centre and edge of the chain.
  In the second set-up (right) the energy is distributed uniformly across each half, with a discontinuity in the energy density at the centre of the chain (indicated by the dotted lines) and the edge.
}
\label{fig:0}
\end{figure}
\section{Model}
The Hamiltonian of what is by now the standard model to study many-body localization-delocalization transition \cite{Luitz:15,Pal:10,DeLuca:13a} is given by
\begin{equation} \label{eq:heisenbergchain}
  H =\sum^{L}_{i=1} \big( J\vec{s}_i\cdot\vec{s}_{i+1}+h_i s^z_i \big) \equiv \sum^{L}_{i=1} H_i,
\end{equation}
where the $h_{i}$ represent static fields on each site $i$ uniformly distributed in the interval $[-h,h]$ and the spins $\vec{s}_i$ are spin-1/2 representations of the $SU(2)$ 
algebra. Periodic boundary conditions are understood throughout.
In what follows we set $J=1$ and denote the Hilbert space size by $\mathcal{N}_H$. 
We recall that numerical work shows a transition to a fully many-body localized phase when $h$ exceeds $h_c=3.7\pm 0.1$ \cite{Pal:10, Luitz:15} 
although this number might be larger \cite{Singh:15}.
For all values of the model parameters, the model conserves the total spin $S_z$ along the $z$ direction; in the rest of the work we have chosen the subspace with $S_z=0$. 
This model is equivalent (through Jordan-Wigner transformation) to that of spinless fermions with nearest-neighbour density-density interactions hopping on a lattice (the $S_z=0$ subspace 
corresponds to half filling).

In this work we are interested in the dynamics of a specially constructed out-of-equilibrium initial state which has an inhomogeneous energy density. 
We are motivated by the primary question of whether ``hot'' and ``cold'' regions in a given isolated quantum system can effectively act as a 
bath for itself, resulting in thermalization. In particular, we study how well a hydrodynamic description of the system fares. To this end we employ two protocols: \emph{a)} we build a close-to-infinite temperature density matrix with a sine-wave energy lump in the energy density profile. This follows the 
technique used in Ref. \cite{Pal:10} to create a magnetization imbalance; and \emph{b)} we construct a pure initial state with a macroscopic energy imbalance and observe its equilibration. From now on, we will frequently refer to \emph{a)} and \emph{b)} above as ``first'' and ``second'' protocol or set-up, respectively.

Both techniques have their advantages and disadvantages. In the sine-wave energy density profile the system is in linear-response regime and we can control the wave number of the initial perturbation $k$ 
but we can not treat large system sizes and are limited at $L=16$. This protocol is primarily used as a check on the next protocol where we can go to much larger system sizes; both lead to qualitatively similar results.

In the second set-up the system is in the fully nonlinear response regime and we cannot finely control the initial perturbation but, after a transient, we can find the diffusion coefficient quite 
accurately and we can, using appropriate numerical techniques, simulate systems up to $L=26$.

In the first protocol, \emph{a)}: We will study a sinusoidal energy lump at infinite temperature constructed by starting with a single mode, mixed state density matrix (evolving a density matrix instead of a pure state does not make a difference in the thermalizing phase)
\begin{equation}
 \label{eq: dm2}
 \rho^{s} = \frac{1}{\mathcal{N}_H}\left[\mathds{1} + \epsilon \sum_{i=1}^{L} \sin{\left(\frac{2\pi (i-1)}{L}\right)}s^{z}_is^{z}_{i+1} \right],
\end{equation}
where $\mathcal{N}_H=\binom{L}{L/2}$ is the dimension of the zero magnetization subspace and the wave mode is $k=2\pi/L$. Defining the local energy density
\begin{equation}
H_i=J\vec{s}_i\cdot\vec{s}_{i+1}+h_i s^z_i ,
\end{equation}
we have that $\langle H_i(t=0)\rangle = \textrm{tr}\left(\rho^{\textrm{s}}H_i\right)= J\epsilon \sin{\left(\frac{2\pi (i-1)}{L}\right)}+O(\epsilon/L)$ 
with the amplitude $\epsilon \ll J,4/L$; the initial energy is thus distributed sinusoidally across the chain as can be seen in the left panel of Fig. \ref{fig:0} 
(the correction is small \emph{and} independent of $h$).

For the second protocol, \emph{b)}, we initialise the state by cutting the chain 
defined by Eq.~\eqref{eq:heisenbergchain} into two open half chains by switching off the boundary 
terms $H^{(B)}=J\vec{s}_{L/2}\cdot\vec{s}_{L/2+1}+J\vec{s}_{L}\cdot\vec{s}_{1}$ (see Fig.~\ref{fig:0}). An 
initial Hamiltonian is defined as 
\begin{equation} \label{eq:intialham}
  H^{(0)}=H-H^{(B)}=H^{(L)}\otimes\mathds{1}^{(R)}+\mathds{1}^{(L)}\otimes H^{(R)}
\end{equation}
where $H^{(L)}, H^{(R)}$ denote the Hamiltonians for the left and right halves of the chain. We then choose an initial state which is a tensor
product of the highest energy eigenstate of $H^{(L)}$ and the groundstate of $H^{(R)}$ so that 
\begin{equation} \label{eq:initialstate}
|\Psi_{\textrm{in}}\rangle=|\Psi_{\textrm{es}}^{(L)}\rangle\otimes|\Psi_{\textrm{gs}}^{(R)}\rangle.
\end{equation}
This state is an atypical, infinite temperature configuration of the system. Once we switch on the two boundary terms, the full unitary evolution generated by (\ref{eq:heisenbergchain}) of the initial state \eqref{eq:initialstate} is non-trivial. Assuming ergodic dynamics, it should eventually allow for energy to 
flow from one side to the other one (and thus for the system to thermalize).
These boundary terms perturb the initial energy levels of the energy lump by $\mathcal{O}(1)$, and therefore the relative mean-squared fluctuations in the energy are 
$\langle\Delta H\rangle/\langle H\rangle\propto 1/\sqrt{L}$, thereby vanishing in the thermodynamic limit, demonstrating that most of the eigenstates $|E_i\rangle$ appearing in the expansion 
$|\Psi_{\textrm{in}}\rangle = \sum_i c_i |E_i\rangle$
are close to the middle of the many-body spectrum. \\
For these two high energy initial states, Eq.~(\ref{eq: dm2}) and Eq.~(\ref{eq:initialstate}), the MBL transition point is predicted to be at $h_c =3.7\pm 0.1$ \cite{Luitz:15}. 
%
\section{Methods}
One approach for evaluating the dynamics is to undertake full diagonalisation of the system to compute its eigenvalues and eigenvectors. 
This allows one to reach up to system sizes $L=16$ spins using moderate computing facilities; this approach is utilised for the first set-up where the density matrix describes the initial state. 
We may thus effect the unitary time-evolution through $\rho^{\textrm{s}}(t) = \textrm{e}^{-itH} \rho^{\textrm{s}}(0) \textrm{e}^{itH}$ from the computed eigenvalues $\{E_k\}$ and eigenvectors matrix $V$, and following this time evolution we are interested in the subsequent evolution of the energy density profile:
\begin{equation}
 \label{eq: timeevolution}
 \textrm{tr}\left(\rho^{\textrm{s}}(t) H_i\right) = \sum_{k,k'} \textrm{e}^{-i(E_k - E_{k'})t}\tilde{\rho}_{kk'}\tilde{H}_{i,k'k},
\end{equation}
where tilde denotes $\tilde{O} = V^{T}OV$ (representation of the operator $O$ in the $H$-eigenbasis). Specifically we study the temporal behaviour of the 
energy imbalance defined by 
\begin{equation} \label{eq:imbalance}
 \Delta E(t)= \langle\Psi(t)|H^{(L)}|\Psi(t)\rangle-\langle\Psi(t)|H^{(R)}|\Psi(t)\rangle.
\end{equation}
5000 to 70 disorder realizations were employed to obtain the disorder-averaged energy imbalance for system sizes $L=10 - 16$.

For the second set-up where we have pure states as initial states with two different homogeneous energy densities (square-wave lump) we can do better: 
because we only need the dynamics generated by the full Hamiltonian,
\begin{equation} \label{eq:quench}
  |\Psi(t)\rangle=e^{-itH}|\Psi_{\textrm{in}}\rangle,
\end{equation}
we may employ the technique of Krylov subspaces that avoids full diagonalisation. We thereby demonstrate that this technique can be used to study dynamics 
in system sizes up to $L=26$.   
10000 to 120 disorder realizations were employed to obtain the disorder-averaged energy imbalance for system sizes from $L=10-22$; the representative results displayed for $L=26$ employed only up to 10 samples.

A similar stratagem for matrix exponentiation was first employed to compute transition amplitudes without explicit knowledge of eigenstates \cite{Nauts}. 
The basic idea of Krylov subspace techniques \cite{Expokit} is to approximate the solution of Eq.~\eqref{eq:quench}, i.e.,
$|\Psi(t)\rangle=|\Psi_{in}\rangle-iHt|\Psi_{in}\rangle-\frac{H^2}{2!}t^2|\Psi_{in}\rangle+\dots$, with an optimal polynomial approximation from within the \textit{Krylov subspace} 
$\mathcal{K}_m=\Span\{|\Psi_{in}\rangle,H|\Psi_{in}\rangle,H^2|\Psi_{in}\rangle \cdots,H^{m-1}|\Psi_{in}\rangle\}$. This is obtained by an Arnoldi decomposition of the matrix 
$A_m = V^T_mHV_m$, where $m$ is the dimension of the subspace ($m \ll \mathcal{N}_H$), $A_m$ is a Hessenberg matrix which is a projection of $H$ onto $\mathcal{K}_m$ with respect 
to the orthonormal basis $V_m$ \cite{Expokit}. The solution is then given by $|\Psi(t)\rangle \approx V_m\textrm{exp}(-itA_m)\ket{e_1}$, where $\ket{e_1}$ is the first unit vector in the Krylov subspace.
The more compact, ($m\times m$ instead of $\mathcal{N}_H\times\mathcal{N}_H$) and projected Hamiltonian $A_m$ is then exponentiated using standard Pad\'{e} techniques \cite{Armadillo}. 
On the other hand, simply summing up the power series of the exponential can yield unstable and inaccurate results unless the number of terms in the series and the machine precision are increased \cite{Loan}.

\section{Transport and diffusion}

\begin{figure*}[ttp]
\centering
  \begin{tabular}{@{}ccc@{}}
\hspace*{0.0cm}\raisebox{5.5ex}{\includegraphics[scale=0.5]{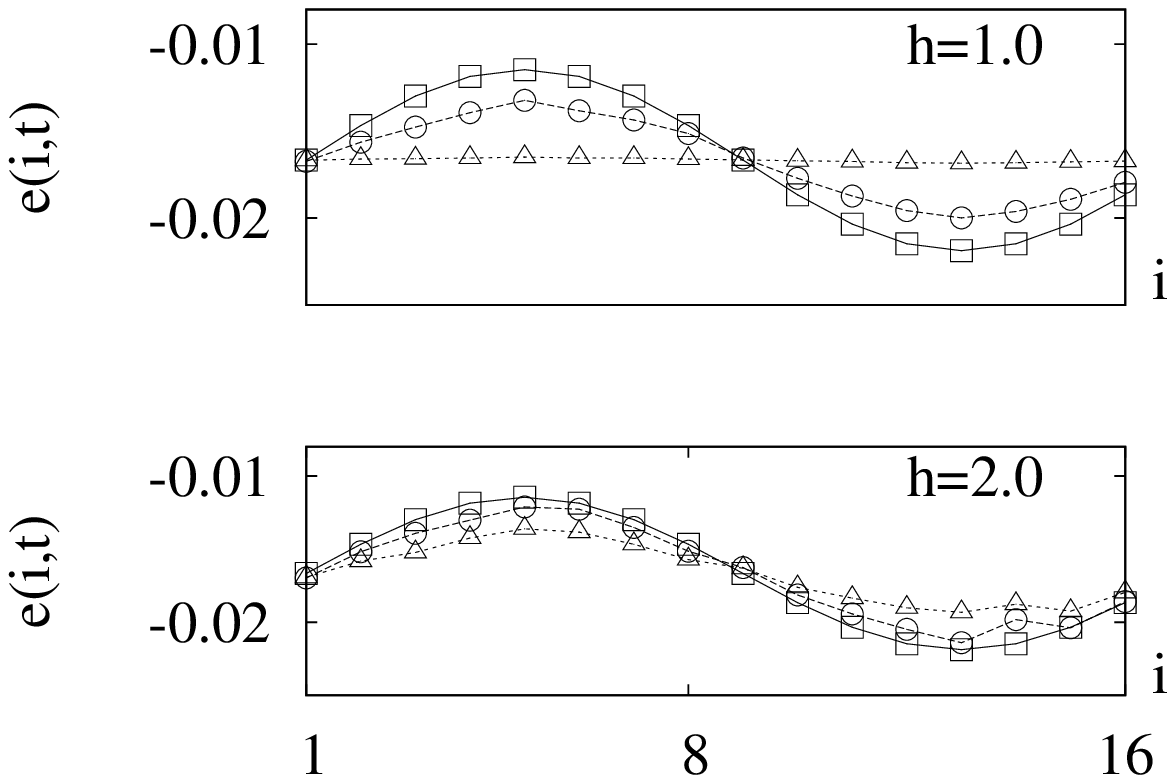}} &
\hspace*{0.0cm}\raisebox{0.5ex}{\includegraphics[scale=0.4]{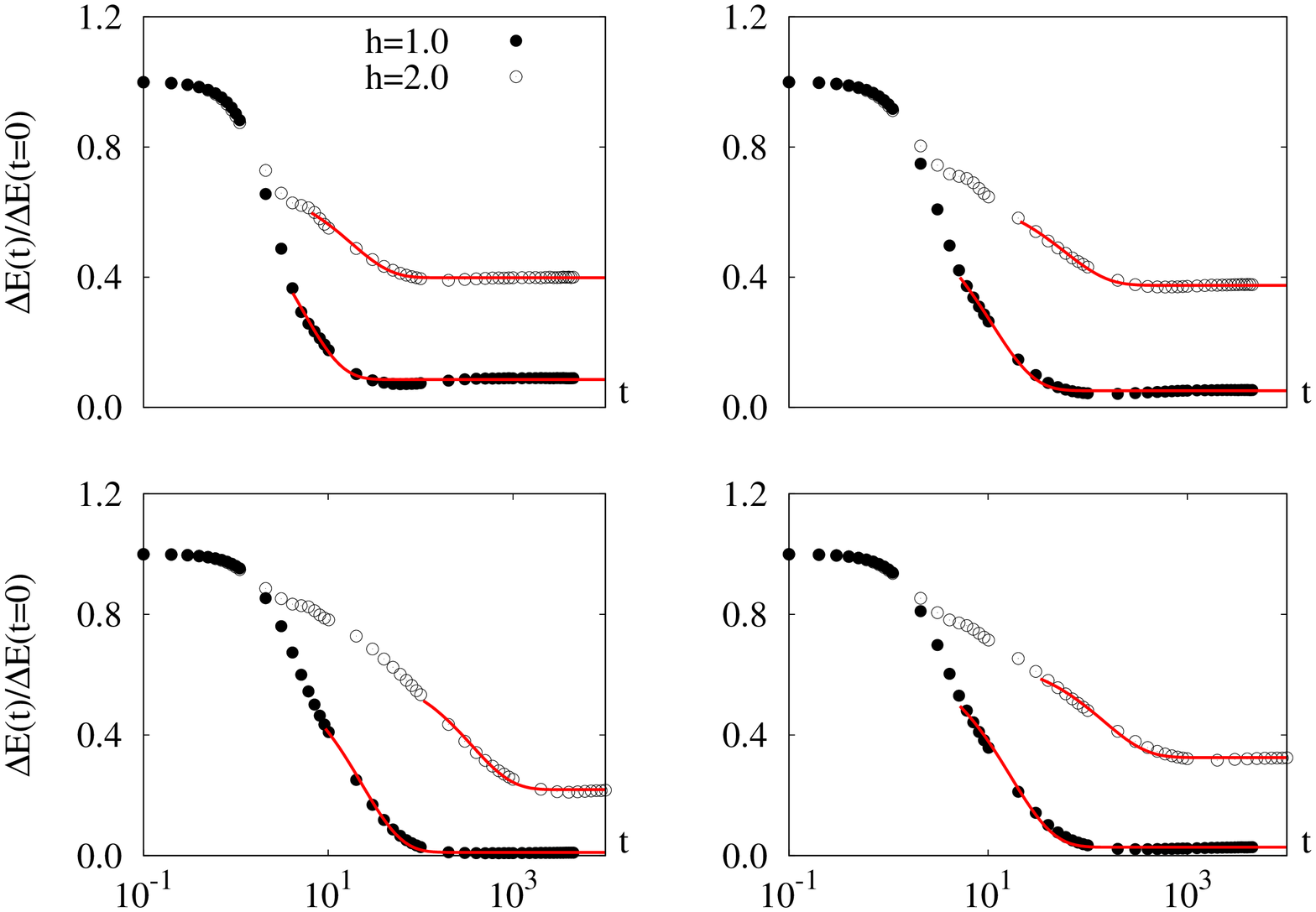}} &
  \end{tabular}
  \vspace*{-0.35cm}
\caption{(Colour online) 
Energy transport in the ergodic phase of the disorder-averaged Heisenberg model with a sinusoidal energy lump for the initial state.
Left top and bottom panels: Spatial energy profiles at various fixed times $t=0, 5, 100$ ($\csquare, \circ, \cvartriangle$) for disorder strengths $h=1$ (top) and $h=2$ (bottom) in 
an $L=16$ chain.
Right 4x4 panels: 
Energy imbalance between the left and right halves of $L=10,12,14,16$ chains (clockwise from top-left) for disorder strengths $h=1, 2$, with the hydrodynamic fits shown as full red lines.
}
\label{fig:1}
\end{figure*}

In the deep localised phase ($h\gtrsim 3.7$) the energy imbalance is seen to persist in the energy density profile evolution over arbitrarily long times, in agreement with the theoretical results on MBL. 
At sufficiently weak disorder intensity, the profile is instead seen to relax to the equilibrium (flat) profile on a finite time scale (see left and top panels of figures \ref{fig:1} and \ref{fig:3}). 
Our primary goal in the following is to establish a quantitative phenomenology for the observed transport.

Hydrodynamics is the macroscopic description of transport. One expects that for disturbances with wavelength $\lambda\gg a$, the lattice spacing, an effective description of transport arises which is 
dependent on the specific microscopic dynamics only through a few transport coefficients, at least for a large class of physically relevant initial states.

The globally conserved quantities in our system are total spin $S_z$ and total energy $E$. The spin density transport in the model under consideration has been investigated in previous works, and has been 
found to be subdiffusive either in the entire ergodic phase \cite{Lev:15, Luitz:15a} or close to the MBL transition \cite{Agarwal}.  
In what follows we will be interested in obtaining an effective hydrodynamic description for the second conserved quantity \textit{viz.} energy density; indeed there is no reason that 
spin and energy transport need occur equally quickly \cite{glasses, Meisner, Gopalakrishnan:15}.
Given a state, pure or mixed, the energy density is given respectively as $e(i,t)=\bra{\Psi(t)}H_{i}\ket{\Psi(t)}$ and $e(i,t)={\rm tr}(\rho(t)H_i).$ The conservation of energy necessitates that $\sum_{i=1}^L e(i,t)=E$ is a constant.

Obtaining a rigorous hydrodynamic framework from an underlying quantum mechanical evolution is by any means a formidable task. Nevertheless in what follows we show that such a 
heuristic approach can indeed work exceptionally well. 
That is, we provide evidence for a simple diffusion law to model the relaxation process to equilibrium in the ergodic regime, but which breaks down well before the value of the disorder strength where the MBL transition is 
expected to occur.

Observation of the relaxation of the energy density profile (see e.g. left and top panels of figures \ref{fig:1} and \ref{fig:3}) suggests a phenomenological diffusion law. Its time evolution should then satisfy the equation
\begin{equation} \label{eq:diffcont}
\frac{\partial{e(i,t)}}{\partial t}={D_e} (\nabla^{2} e(t))(i),
\end{equation}
where $\nabla^2$ is the laplacian on the lattice, which in our case becomes \[(\nabla^2 e(t))(i)=e(i+1,t)-2e(i,t)+e(i-1,t)),\]and  $D_e$ is a diffusion coefficient. 
We expect that this equation may effectively describe the emergent transport behaviour of energy on a large, coarse-grained, spacetime scale.

Let us first focus for definiteness on the first set-up, where the initial state (\ref{eq: dm2}) gives an energy density profile shaped as the longest harmonic,
\begin{equation}
 \label{eq:tmp1}
e(i,0)=A(0) \sin\left(\frac{2\pi (i-1)}{L}\right)
\end{equation}
with $A(0)\propto \epsilon$. If a diffusion equation is satisfied, then the sinusoidal shape of the profile is maintained during time evolution and its amplitude decreases in time as
\begin{equation}
 \label{eq:tmp2}
A(t)=A(0) \,e^{-\gamma t},
\end{equation}
where $\gamma=D_e 4\pi^2/L^2$. 

The observation of the facts that the shape of the profile is maintained during time evolution and that its amplitude is exponentially damped in time, alone, would simply indicate compatibility with a generic translation-invariant phenomenological equation of the form
\begin{equation}
\frac{\partial{e}}{\partial t}=f\big(\nabla^{2}\big) \, e,
\end{equation}
with the function $f$ undetermined. What really indicates diffusive behaviour (i.e., $f(\nabla^{2}) \propto \nabla^{2}$) is the precise scaling of the exponential decay rate $\gamma$ with the system size $L$,
\begin{equation}
\gamma = D_e \frac{4\pi^2}{L^2},
\end{equation}
for some fixed, $L$-independent value of $D_e$. A different scaling law would imply a different phenomenological equation; 
e.g. $\gamma=\tilde{D}_e (2\pi/L)^{2+b}$, with $b>0$, would imply a subdiffusion law (corresponding to $f\big(\nabla^{2}\big) \propto (\nabla^{2})^{1+b/2}$). 
In this light, the crucial step of our analysis will be the determination of the scaling law of the extracted values of $\gamma$ vs $L$.

We stress that we are interested in capturing a diffusive regime in these systems, and interpret the vanishing of the diffusion constant before the MBL transition as the possible onset of subdiffusion. 
This approach, as opposed to directly modelling a fractional transport equation with a generic exponent $b$, is reasonable for many reasons: 
(i) diffusive transport of conserved quantities is generically expected in disordered systems \cite{Anderson:58, Abrahams:10} for some range of disorder strengths, 
in particular also expected and observed in the many-body systems \cite{Basko:06, Agarwal, Znidaric:16};
(ii) at weak disorder strengths for the systems of sizes we are able to access, had we modelled the dynamics with a generic transport equation with a nonzero $b$, 
an erroneous value of the extracted $b$ will model the data \cite{Znidaric:16} because large scattering lengths in this limit can overestimate the transport rate thereby incorrectly 
suggesting anomalous transport \cite{Luitz:15a};
(iii) subdiffusion entails a space-dependent (or equivalently time-dependent, when space and time are nontrivially related $\sqrt{\Delta x^2} \sim t^{\beta}$) 
diffusion constant $D_e (x) \sim x^{2\beta - 1}$ \cite{Agarwal, Znidaric:16}, with $\beta = 1/2$ for diffusion and $\beta < 1/2$ for subdiffusion. 
Therefore when the space/time-dependence of the extracted $D_e(x)$ drops out, we may interpret the transport as being normal diffusive; whereas when the 
space/time-dependence of $D_e(x)$ continues to the thermodynamic limit, we may interpret the transport as being subdiffusive. 
This is precisely what our extrapolations of $D_e(L)$ to the thermodynamic limit achieves.

Our analysis thus consists of, for each value of the disorder strength $h$, i) observing such an exponential decay, ii) extracting the decay constant $\gamma(L)$, and from this iii) 
the scaling vs $L$ of the quantity $D_e(L) := \gamma(L) \, L^2 /(4\pi^2)$. 
If $D_e(L)$ settles to a finite nonzero value as larger and larger system sizes are considered, 
we can claim that the observed transport is compatible with a diffusion equation with $D_e \equiv D_e(L\to\infty) > 0$ \cite{NoteArg}. 
This is indeed what we find, as explained in the next section.
Note that when the system thermalizes, whether by 
diffusion or subdiffusion, the (sub)diffusion coefficient $D_e(L)$ is dependent only on the system size $L$ at long-times i.e. there is no characteristic time-scale involved apart from that defined by 
$\gamma(L)^{-1}$. 
This is analogous to the one-particle problem where the Thouless time-scale $\tau_T \propto L^2/D$ captures the essence of diffusion for a \textit{fixed} system size $L$ \cite{EdThouless}.

Instead of the amplitude $A$ one can use the energy imbalance defined by Eq~\eqref{eq:imbalance} that can be written as $\Delta E(t)=\sum^{\frac{L}{2}}_{i=1}e(i,t) -\sum^{L}_{i=\frac{L}{2}+1} e(i,t)$. 
This quantity behaves in a similar way but is easier to extract.

Let us now focus on the second protocol. The initial state defined by Eq.~\eqref{eq:initialstate} will exhibit an approximate square-wave shape in the energy density profile; 
it should be rigorously so in the thermodynamic limit and sufficient number of disorder averaging. 
As it contains many modes, the shape is not retained during time evolution; see top panel of Fig. \ref{fig:3} where this fact is visible, though masked by disorder fluctuations. 
Nevertheless, assuming the transport equation Eq. \eqref{eq:diffcont} applies, by solving it one can straightforwardly derive the following expected evolution of the energy imbalance for a periodic chain, 
\begin{equation} \label{eq:diffsol}
\Delta E(t)=\Delta E(0)\frac{8}{\pi^2}\sum_{n\in 2\mathbb{N}+1}\frac{e^{-{D_e}(\frac{2\pi n}{L})^2t}}{n^2}.
\end{equation}
The procedure to establish a diffusive behaviour is then akin to that pertaining to the first protocol above: for fixed $h$ we extract the parameter $D_e(L)$ 
by fitting the numerical data for $\Delta E(t)$ with the functional form of Eq. \eqref{eq:diffsol}, after appropriately truncating away a 
transient regime for the different system sizes and disorder strengths. The truncation is implemented such that (i) the extracted $\gamma(L)$ are relatively stable with respect to the time of truncation, 
and (ii) the maximum time-range is reasonably captured by the fit \cite{NoteArg2}.
If $D_e(L)$ settles to a finite value as $L\to\infty$, we may claim compatibility with the diffusion law.

For the extrapolation we find that an exponential function $D_{e}(L) = D_{e} + c_0\exp(-c_1 L)$ is a good fit to the data as long as $D_e>0$ measurably. 
In contrast, the absence of diffusion or the presence of subdiffusion would be signalled by 
$\lim_{L \rightarrow \infty} D_e(L)=0$ [such that $D_e(L)\simeq D_e L^{-b}$ for subdiffusion].

\section{Analysis and results}
\textit{Sinusoidal wave}: We first study the energy diffusion with initial condition given by a mixed state 
Eq. \eqref{eq: dm2} through an analysis of the energy imbalance between the left and right halves of the chain, as explained in the previous section. 
Note that the single-mode initial condition is independent of sample and disorder strength for a given system size $L$ because the 
$z$-field term does not contribute to the energy density: $\textrm{tr}(\rho^{\textrm{s}}s^z_i) = 0$.

The left panel of of Fig. \ref{fig:1} displays the energy profiles at various fixed times for these disorder strengths, illustrating how the sinusoidal shape of the 
$k$-mode is well retained. The disorder-averaged results at disorder strength $h=1, 2$ for $L=10,12,14,16$ chains (clockwise from top-left) are displayed in the right 4x4 panels.
The full red line denotes a fit to an exponential decay, given by the diffusion law,
\begin{equation}
 \label{eq: fullfit}
 \Delta E(t)/\Delta E(0) = c_{\infty} + b\exp{(-\gamma t)},
\end{equation}
where the free parameters $c_{\infty},b,\gamma$ are extracted for each dataset $(h,L)$. For all of them the functional form \eqref{eq: fullfit} fits rather well. 
The offset $c_{\infty}$ is to account for finite-size effects, and is seen to vanish as $L\to\infty$ in the region compatible with diffusion.
 \begin{figure}[ttp]
\centering
\hspace*{0.5cm}\raisebox{0.5ex}{\includegraphics[scale=0.65]{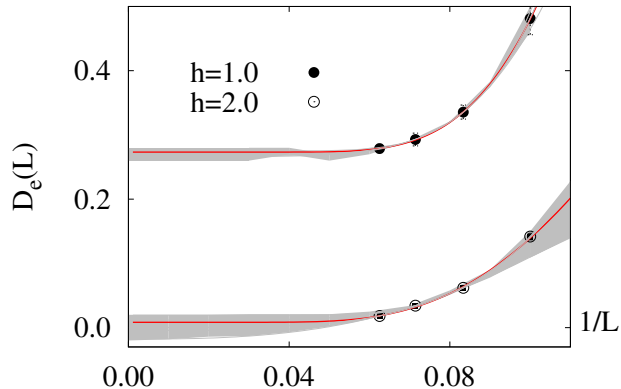}} 
  \vspace*{-0.35cm}
\caption{(Colour online) 
 Scaling of diffusion constant $D_e(L) := L^2\gamma/4\pi^2$ with inverse system size $1/L$ at disorder strengths with the sinusoidal energy lump as initial condition; 
  extrapolation with an exponential scaling is shown. Note the qualitative and quantitative difference in the extrapolated values for the two cases, suggesting 
  considerably weaker diffusion (arguably even subdiffusion) for $h=2$ as compared to $h=1$ which has a clear nonzero thermodynamic value for $D_e(L)$.
  The grey lines show the uncertainty in the fits from a stability analysis (see text).
}
\label{fig:2}
\end{figure}
 We scale the extracted diffusion constants $D_e(L) = L^2\gamma(L)/4\pi^2$ with the inverse system size for a range of system sizes.
 As displayed in Fig. \ref{fig:2} we exponentially extrapolate to the thermodynamic limit.
 For analyzing the stability of our fits and extrapolations we consider both exponential and polynomial fits, as well as fitting only a certain range of the data points;
 the uncertainty in the fits are denoted by the grey shaded area in Fig. \ref{fig:2}.
 A clear discernible difference is apparent between the two cases: 
for $h=1$, $D_e(L\to\infty)$ is finite whereas for $h=2$, $D_e(L\to\infty) \approx 0$. 
This suggests diffusion in the former case and drastically suppressed diffusion (or arguably even subdiffusion in the latter, where we find 
a power law extrapolation with zero offset works well too \cite{NoteArg}). 
This is confirmed by a more thorough analysis of the second (square-wave) protocol, where we may go up to much larger system sizes using the Krylov technique on pure states. 

\begin{figure*}[http]
\centering
  \begin{tabular}{@{}cccc@{}}
      \hspace*{3.51cm}\raisebox{-25.ex}{\includegraphics[angle=270, origin=c, scale=0.36]{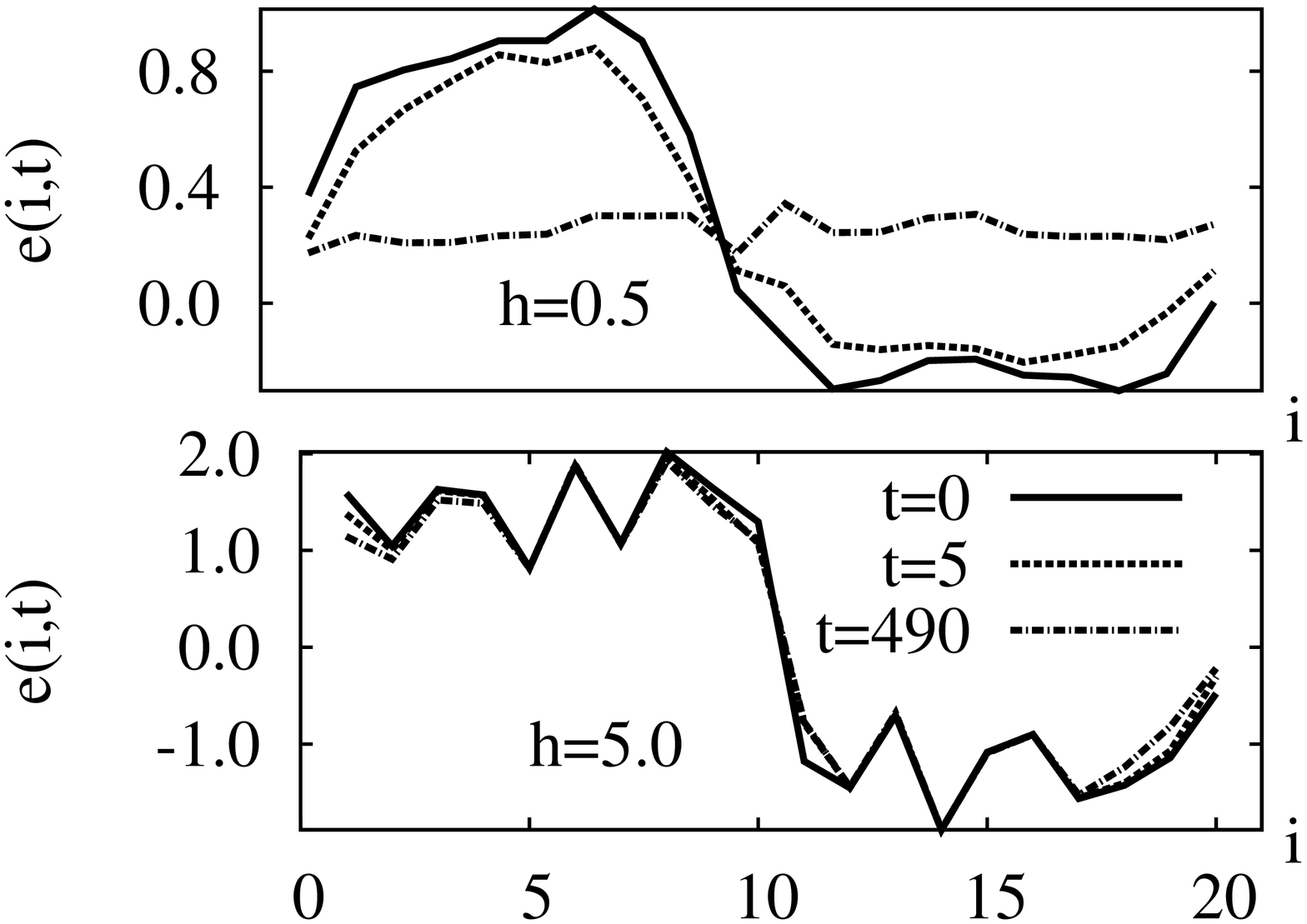}} \\
      \hspace*{-4.01cm}\raisebox{-2ex}{\includegraphics[scale=0.6]{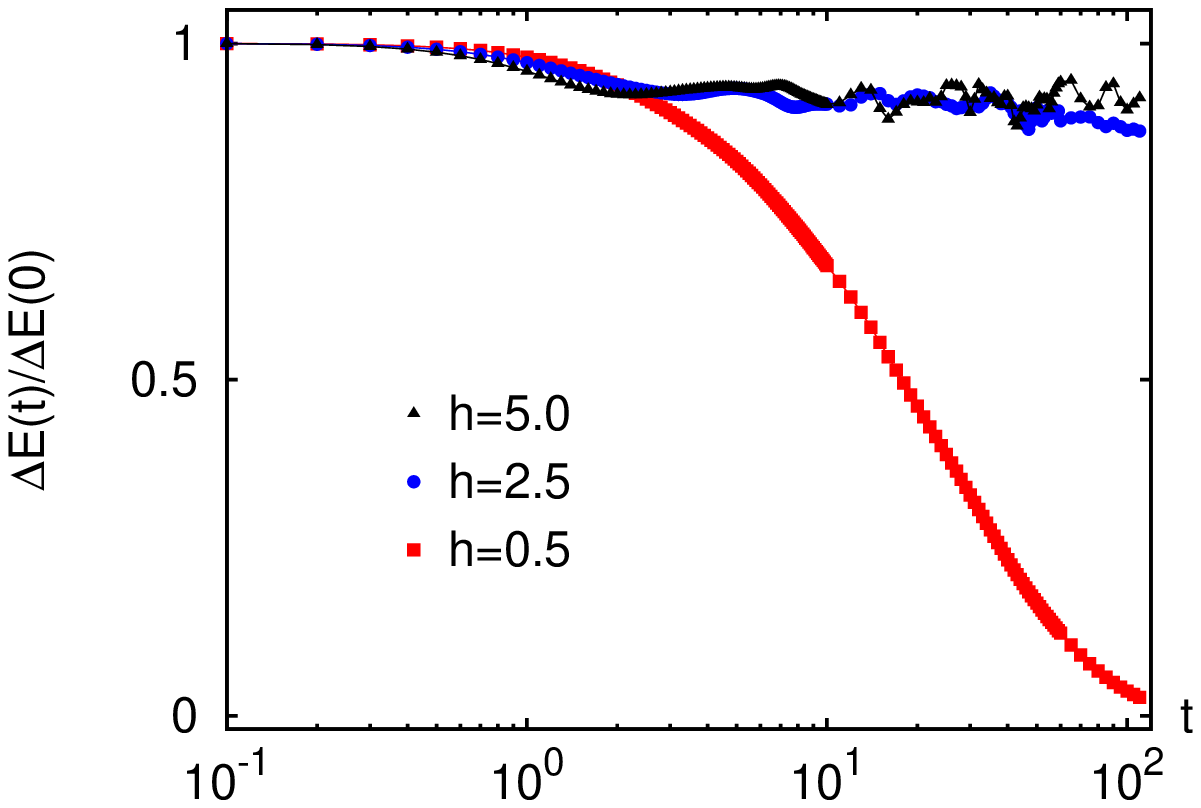}} &
      \hspace*{-5.01cm}\raisebox{-2ex}{\includegraphics[scale=0.6]{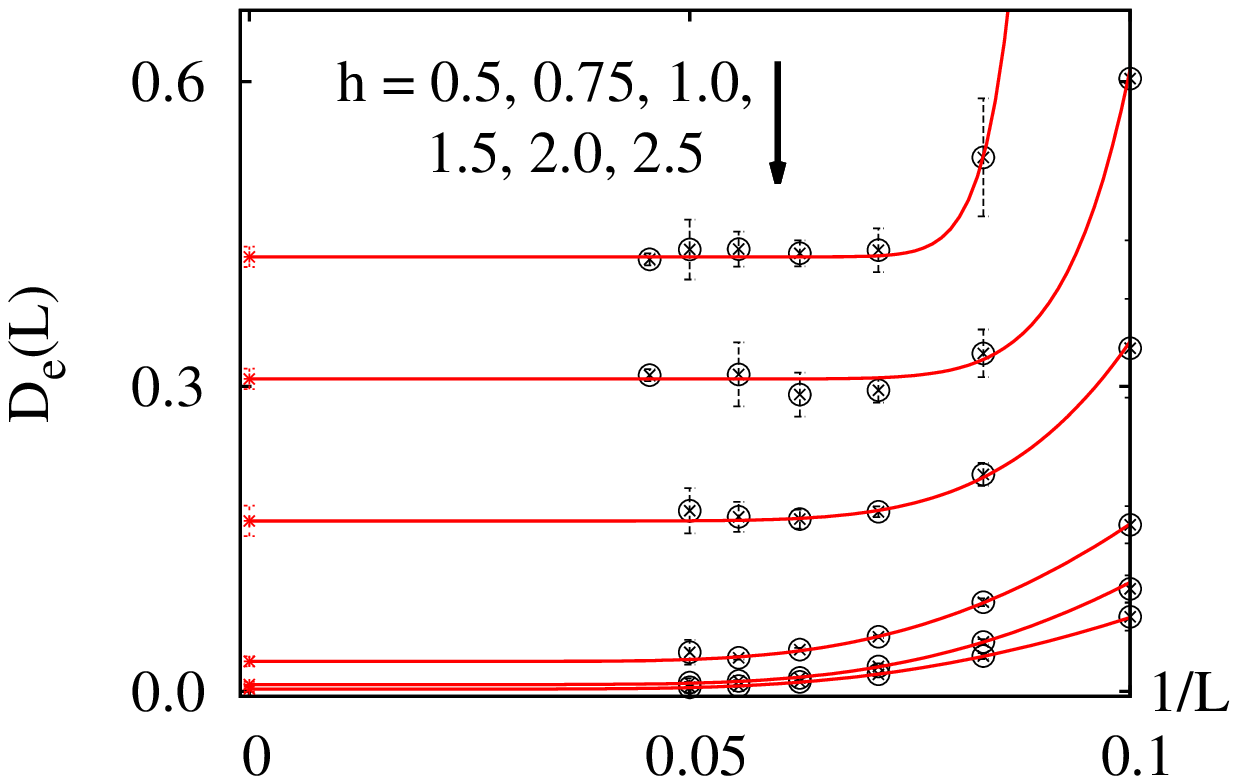}} 
  \end{tabular}
  \vspace*{0.25cm}
  \caption{(Colour online) Energy diffusion in the ergodic phase of the disordered Heisenberg model within the second protocol (square-wave initial state). 
Top panel: Disorder-averaged spatial profile of local energies $e(i,t) \equiv \langle H_i \rangle_t$ for a periodic $L=20$ chain at disorder strengths $h=0.5$ (top) and $h=5.0$ (bottom), 
at various fixed times $t=0, 5, 490$. Note the energy equilibration in the weak disorder regime at long times but the absence of any transport deep in the MBL phase. 
Left panel: Disorder-averaged energy imbalance for disorder strengths $h=0.5,2.5,5$ for which the disordered spin chain is expected to be deep in the ergodic phase, close to the localisation 
transition, and well within the MBL phase respectively for system size $L=26$. 
Right panel: 
Exponential extrapolation functions (solid lines) of $D_{e}(L) = D_{e} + c_0\exp(-c_1 L)$ to the thermodynamic limit at various fixed disorder strengths, similar to Fig. \ref{fig:2}.
The $3\sigma$ confidence intervals of the extracted $D_e(L)$ are indicated along with that in the $L\rightarrow \infty$ extrapolated values.
}
\label{fig:3}
\end{figure*}
\textit{Square wave}: We now turn to the second protocol, with a square-wave initial profile, in order to sharpen and corroborate the findings from the analysis of the single mode as initial state; moreover here we can 
treat larger system sizes as explained previously. 
Let us first draw a broadbrush picture from the spatial energy density profiles as the system evolves. In the top panel of Fig.~\ref{fig:3} we display the energy profile 
across a chain of length $L=20$ at three different fixed times in the weak (top plot) and strong (bottom plot) 
disorder regimes. In the former case we see that there is energy equilibration at long times and the entire chain reaches a uniform energy density. 
This is to be contrasted with the strong disorder case where the spin chain is expected to be in the MBL phase: no transfer of energy is observed for a wide range of times 
and no effective temperature may be defined for the system \cite{hanggi}. 

The same qualitative picture might also be inferred from the time evolution of the energy imbalance between left and right halves of the chain.
The time evolution of the disorder-averaged energy imbalance \eqref{eq:imbalance} for a larger 
system size $L=26$ are plotted in the left panel of Fig. \ref{fig:3} for three regimes: (a) weak disorder limit $h=0.5$, (b) close to the transition $h=2.5$, and (c) deep in the 
localised phase $h=5$ (recall that the critical value for the MBL transition is $h_c=3.7\pm 0.1$). 
The qualitative behaviour is as follows.
In the first case, $h=0.5$, $\Delta E$ goes quickly to zero at long times, to a situation of a uniform energy density across the entire chain.
In the third case $h=5$, $\Delta E$ clearly does not decay in time. This is due to the lack of energy transfer which is expected in the MBL region, where there is no equilibration.
In between the two cases, for $h=2.5$ for example, there is a very slow relaxation process governing the dynamics of the system which 
is very weakly diffusive or even possibly subdiffusive with a continuously changing dynamical exponent. 

\begin{figure}[bbp]
\centering
\hspace*{-0.5cm}\raisebox{0.5ex}{\includegraphics[scale=0.5]{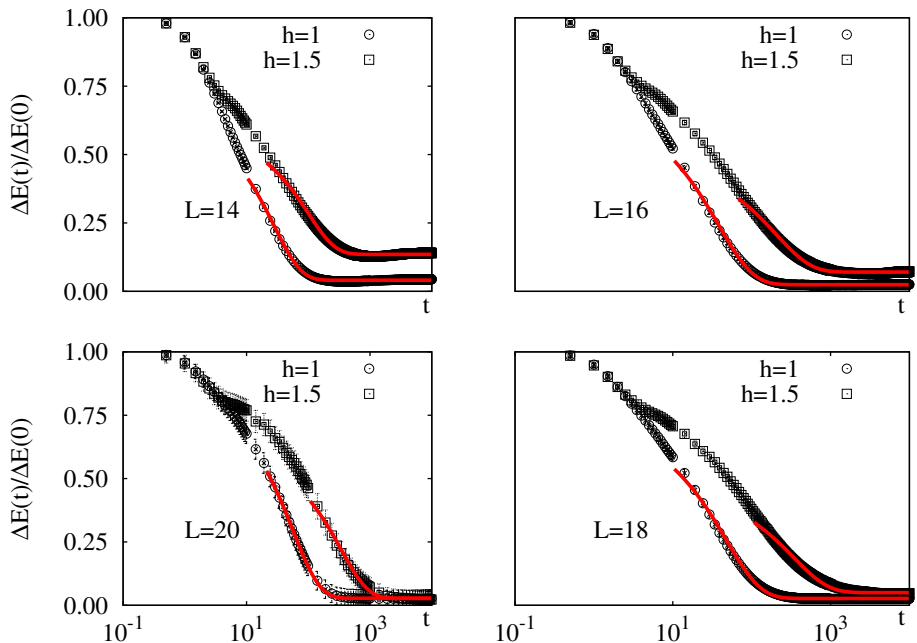}} 
  \vspace*{-0.35cm}
\caption{(Colour online) 
Disorder-averaged energy imbalance as a function of time with a square wave energy density as the initial state for a range of system sizes $L$ and disorder strengths $h$. 
Hydrodynamic diffusion fits using Eq. \eqref{eq:finalfit}, up to $n=11$, for a range of system sizes and disorder strengths. This procedure was employed to extract the diffusion constants plotted in 
the right panel of Fig. \ref{fig:3}.
}
\label{fig:fits}
\end{figure}

We may perform a similar analysis, \textit{mutatis mutandis}, for the present case as was done for the sinusoidal wave in the right 4x4 panels of Fig. \ref{fig:1}. 
In this case, the analysis was performed for a larger range of disorder strengths and system sizes, which helps to substantiate our previous claims: for $h=0.5,0.75,1,1.5,2,2.5$ and $L=12,14,16,18,20,22$,
temporal diffusion fits for $\Delta E(t)/\Delta E(0)$ are undertaken, with a modification of the functional form \eqref{eq:diffsol},
\begin{equation}
\label{eq:finalfit}
\Delta E(t)/\Delta E(0)=c_{\infty}+A_0 \sum_{n\in 2\mathbb{N}+1}\frac{e^{-{D_e}(\frac{2\pi n}{L})^2t}}{n^2}.
\end{equation}
The free fit parameters are $c_{\infty}, A_0, D_e$. 
Some of these diffusion fits for a couple of $h$ and $L$ values are shown in Fig. \ref{fig:fits}.
Just as before for the sinusoidal case and the fit Eq. \eqref{eq: fullfit}, the functional form given by the diffusion law fits Eq. \eqref{eq:finalfit} quite well all the datasets for the square wave case; 
it turns out that only the first few terms give a substantial contribution to the above series in the fit (we used upto $n=5$). 
The offset $c_{\infty}$ is to account for finite-size effects, and is seen to vanish as $L\to\infty$ in the region compatible with diffusion. 
A short-time parabolic transient is present in the numerical curves for the energy imbalance, which is a quantum-mechanical effect that has nothing to do with transport 
and must be appropriately truncated away; the truncation is implemented in such a way that i) the extracted $\gamma(L)$ are relatively stable with respect to the time of truncation, 
and ii) the maximum time-range is reasonably captured by the fit \cite{NoteArg2}.

For each given disorder strength $h$ the extracted values $D_{e}(L)$ for various system sizes $L$ allow for extrapolation to the thermodynamic limit. 
We find that an exponential function is a good fit to the data as long as $D_e>0$ measurably: 
the fit parameters $D_e(L)$ are extracted and are scaled with $L$ as shown in the right panel of Fig. \ref{fig:3} for various fixed disorder strengths $h$. 
The exponential infinite-size extrapolation fits are displayed by solid red lines at each disorder strength as a function of inverse system size, limiting to a finite value for weak disorder strengths.
In contrast, since only the first harmonic gives substantial contribution, a subdiffusion law would be signalled by 
$\lim_{L \rightarrow \infty} D_e(L)=0$ and a power law fit for finite-size corrections $D_e(L)\simeq \tilde{D}_e L^{-b}$, where the exponent $b>0$ is the subdiffusion exponent and 
$\tilde{D}_e$ is to be interpreted as the subdiffusion coefficient.

\begin{figure}[ttp]
\centering
\hspace*{0.5cm}\raisebox{0.5ex}{\includegraphics[scale=0.7]{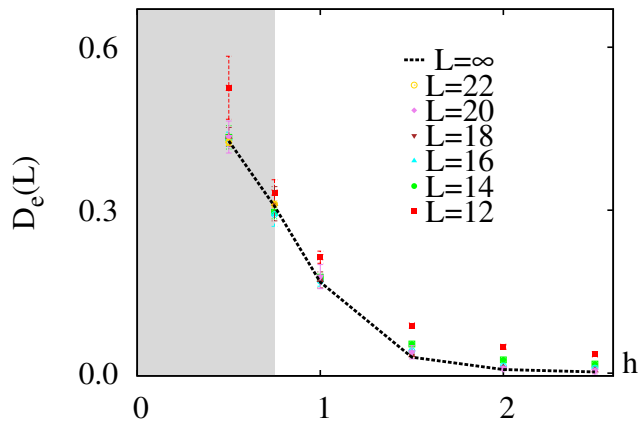}} 
  \vspace*{-0.35cm}
\caption{(Colour online) 
Diffusion constant as a function of disorder strength in the system with the square wave initial energy lump 
(showing here the same values of bottom-right panel in Fig. \ref{fig:3} as a function of $h$). The diffusion constant vanishes well before the localization transition, which 
occurs at $h_c \approx 3.7$; the dashed line is the thermodynamic result obtained from bottom-right panel in Fig. \ref{fig:3} and may be taken as a guide to eye. 
The shaded area indicates a regime where transport rates might be overestimated due to the rarity of impurity scattering events leading to a remnant of quasi-clean transport, 
as explained in the text \cite{Znidaric:16}.
}
\label{fig:4}
\end{figure}

That the extrapolated ($L\to\infty$) diffusion coefficients are finite for sufficiently weak disorder substantiates the claim that the model displays diffusive transport in this range of parameters. 
The value of $D_e$ decrease as the disorder $h$ is increased and becomes compatible with zero for $h=h^*\gtrsim 2$, implying a breakdown of diffusion around this value of disorder strength and the 
possible onset of subdiffusive energy transfer processes.
The dependence of the values $D_e(L)$ extracted from the fit and of their extrapolation to infinite size 
$\lim_{L \rightarrow \infty} D_e(L)$ on the disorder strength is plotted in Fig. \ref{fig:4}. The same point as with the sinusoidal lump is more clearly apparent here: diffusion is suppressed 
around $h=h^*\gtrsim 2$, well before the onset of full many-body localization. 
This encapsulates the two main findings of this work: that there is a energy diffusion in the ergodic phase and that it 
ceases well before the localization transtion.

We mention a caveat at this stage: For weak disorder strengths there is the possibility that on short length and time scales 
accessed here the rarity of scattering processes might lead to an overestimation of the transport rate; for spin diffusion such a length scale is $L_* \sim 1/h^{\nu}$, with $\nu \approx 1$ \cite{Znidaric:16}. 
Such a critical length scale must arguably hold here too for energy transport; clearly this effect is relevant for our studied system sizes only for $h/J \ll 1$ and not when $h/J \sim O(1)$. Nevertheless 
we indicate in Fig. \ref{fig:4} a shaded area where our transport rate might have been overestimated. 
The value $h^*\approx2$ is thus the result of the best available numerics for the onset of the Griffiths effects which are the cause of subdiffusion. One should however not forget the finite size effects
which become strong approaching the MBL transition; the transport phenomenology emerges at length scales bigger than any disorder-born correlation length, which was identified to exist and be less than, 
but still of the order of, the system sizes used in the present work at $h=2.5$~\cite{pietracaprina:16}.

We also note that the extrapolation of the $D_e(L)$ values obtained for the sinusoidal and square wave protocols are not in agreement within the statistical errors 
[at $h=1$ $\lim_{L\rightarrow \infty} D^{\textrm{sqr.}}_e (L) \approx 0.6 D^{\textrm{sin.}}_e (L)$]. This can be due to an underestimation of $D_e(L)$ for the square-wave case from the higher modes still being present there, or 
it could be due to an inaccurate extrapolation of 
the results from the small system sizes studied in the sinusoidal case (recall that for the single-mode case the largest system size is $L=16$, while for the square-wave it is $L=22$), or both.
Nevertheless there is good qualitative agreement between the two cases vis-\'{a}-vis a large part of the ergodic phase harbouring substantial energy diffusion, and another large part of the same phase 
where energy diffusion is suppressed (and arguably becomes subdiffusive).

\section{Discussions and conclusions}
In summary, we have studied the quantum dynamics of inhomogeneous energy density profiles in a disordered Heisenberg spin chain by means of a numerical Krylov subspace technique. Our results indicate that 
(i) energy transport is diffusive in an extended region of the delocalized phase, and (ii) is transported either with exponentially suppressed diffusion or arguably even subdiffusively 
well before the transition to the localised phase. Although it is possible that we have overestimated transport rates in the diffusive regime due to the accessible time-scales being short, the goodness of the 
diffusion fits \textit{and} the scaling of the extracted $D_e(L)$ constants behove us to suggest otherwise. Moreover the overestimation of transport rate is a severe possibility only when $L \ll L_* \sim 1/h^\nu$, 
with $\nu > 1$, due to rare scattering processes over small lengths $L$, leading to quasi-clean transport behaviour \cite{Znidaric:16}; for $h/J = \mathcal{O}(1)$ such an issue is no longer pertinent.  

The first of our findings $-$ the observation of energy diffusion breaking down at $h^* \approx 2$, which is approximately the point at which a recent study
\cite{Luitz:15a} find the spreading of entanglement entropy to change from diffusive to subdiffusive $-$ further points towards a more exotic ergodic phase which harbours two subphases 
with vastly differing dynamics of energy transport. 
The nondiffusive energy transport which occurs between $h^*\approx2$ and $h_c=3.7$ should stop altogether beyond the MBL transition. A similar diffusive-subdiffusive transition for the spin dynamics 
within the ergodic phase has been established in a number of works already \cite{Agarwal, Znidaric:16}.

The second of our findings agrees qualitatively with findings from other works \cite{potterprx, voskprx} (where deep in the ergodic phase energy is expected to be diffusive)
but is in contrast with some of the existing literature \cite{Luitz:15a} (where anomalous transport of entanglement entropy is expected to occur 
throughout the entire ergodic phase). 
This indicates that the existing phenomenology of the Griffiths effects which accounts for transport in these systems needs to accommodate for the diffusion of one conserved quantity 
in the ergodic phase without it simultaneously aiding an \textit{equally fast} thermalization of another conserved quantity. In fact after our work was completed, 
Ref. \cite{Gopalakrishnan:15} was published, which contains an improved analysis of the renormalization group results and can accommodate the observations of our paper i.e. 
the coexistence of anomalous and normal diffusion in the ergodic phase. 

We have shown that a refined phenomenology of the Heisenberg model with disorder is needed (and possibly of other many-body localizable models) including the possibility of a separate fully ergodic, 
diffusive phase and a subdiffusive phase with respect to energy transport.

\section*{Acknowledgements}

The authors are grateful to I.~Girotto for help with the Argo computing facilities at ICTP. AS would like to thank the Physics Department of the University of Bari, 
where part of this project has been completed, for hospitality and financial support; VKV acknowledges the hospitality received at the Kavli Institute for Theoretical Physics 
where part of this project has been completed; FP and AL would like to thank ICTP for hospitality.
We would also like to thank J. Eisert, C. Gogolin, P. Facchi, R. Fazio, D. Huse, V. Oganesyan, A. Pal, M. \v{Z}nidari\v{c}, M. Schiulaz and R. Vasseur for discussions.


\begin{thebibliography}{100}
\bibitem{NoteArg} A numerically vanishing $D_e(L)$ might also arise from an exponentially suppressed $D_e$ rather than a finite $b$, which is still consistent with the first point of our paper  
i.e. there is energy diffusion in a large part of the ergodic phase.
\bibitem{NoteArg2} The lesser the time-range that is included in the fit (for instance, just the extreme tails) the more will $\gamma(L)$, and thence $D_e(L)$, be suppressed;
it is then only sensible that we take the maximum time-range for which the exponential decay models the data.
\end{thebibliography}
\end{document}